\documentclass[pre,aps,twocolumn,psfig,preprintnumbers,amsmath,amssymb,showpacs]{revtex4}
\usepackage{graphicx}\usepackage{epsfig}%\usepackage{portland}\usepackage{graphics}
\begin{document}
%\preprint{preprintnumber}
\title{Extended Gibbs ensembles with flow}
%
%   --- AUTHORS
%
\author{M. J. Ison$^{1,2}$\footnote{Present address: Department of Engineering, 
University of Leicester, LE1 7RH, UK}} 
\author{F. Gulminelli$^{2}$\footnote{member of the Institut Universitaire de France}}
\author{C. Dorso$^{1}$}
%\affiliation{$^{1}$~your affiliation}
\affiliation{$^{1}$~Departamento de F\'{\i}sica, Facultad de Ciencias Exactas y Naturales,
Universidad de Buenos Aires, Buenos Aires (1428), Argentina}  
\affiliation{$^{2}$~LPC Caen, ENSICAEN, Universit\'e de Caen, CNRS/IN2P3, Caen, France}
%
%   --- DATE
%
\date{\today}% It is always \today, today,
%
%   --- ABSTRACT
%
\begin{abstract}
A statistical treatment of finite unbound systems in the presence of collective motions
is presented and applied to a classical Lennard-Jones Hamiltonian, numerically simulated
through molecular dynamics. In the ideal gas limit, the flow dynamics
can be exactly re-casted into effective time-dependent Lagrange parameters acting on a standard Gibbs 
ensemble with an extra total energy conservation constraint. Using this same ansatz 
for the low density freeze-out configurations of an interacting expanding system, 
we show that the presence of flow can have a sizeable effect on the microstate distribution. 
\end{abstract}
%
%   --- PACS
%
\pacs{64.10.+h, 25.75.Ld, 05.10.-a}  
%
%\pacs{68.35.Rh}{Phase transitions and critical phenomena}
%\pacs{51.30.+i}{Thermodynamic properties, equations of state}
% 64.60.-i General studies of phase transitions
% 05.70.Fh Phase transitions: general studies
% 05.70.-a Thermodynamics
% 51.30.+i Thermodynamic properties, equations of state
 \maketitle
%
%%%%%%%%%%%%%%%%%%%%%%%%%%%%%%%%%%%%%%%%%%%%%%%%%%%%%%%%%%%%%%%%%%%
%       INTRODUCTION:  
%%%%%%%%%%%%%%%%%%%%%%%%%%%%%%%%%%%%%%%%%%%%%%%%%%%%%%%%%%%%%%%%%%%
%
\section{INTRODUCTION}
The thermodynamics of all isolated finite unbound systems is characterized by 
an irreducible time dependence. 
In condensed matter physics, clusters dissociation induced by photo-ionization
 \cite{Catherine,Haberland,Brockhaus,Martinet}
or charge transfer collisions \cite{Farizon,Gluch}, cannot be studied without properly accounting
for the time window of the experiment \cite{Catherine}.
Going down to the femto-scale, the thermodynamic properties of nuclear systems  
can only be accessed through collisions \cite{Wci}.
In the cluster world the dynamical evaporation may still be associated to the thermodynamics of the 
liquid-vapor phase transition \cite{Catherine,Haberland,Farizon} making use 
of a time-dependent temperature within the concept of an evaporative ensemble \cite{Klots,Calvo};
however in the nuclear collisions case the time scales can be so short that the reaction 
and decay channels cannot be decoupled, collective flows appear, and the statistical 
equipartition hypothesis breaks down \cite{Fopi}. If in the Fermi energy regime 
and in the associated multi-fragmentation phase transition these collective flows may be 
only a perturbation in the global energetics, this is not true at SIS energies where 
they are likely to influence light cluster formation by coalescence \cite{Mishustin}. 
In the ultra-relativistic regime the ordered and disordered motions become comparable in magnitude
 \cite{shm}, and collective flows are believed to play an essential role in the characteristics of the 
transition to the quark-gluon plasma observed in the RHIC data \cite{Shuryak,Ko,hydro}. In particular, 
correlations and recombination of thermalized quarks from a collectively flowing deconfined quark plasma, 
is supposed to be the dominant mechanism for soft-hadrons production \cite{Shuryak,Fries}.

In all these very different physical situations, the huge number of available channels and the 
general complexity of the systems under study clearly calls for a statistical treatment.
However, the irreducible time dependence of the process makes the definition of statistical
concepts like statistical ensemble, temperature, pressure, etc. unclear \cite{Koch,Becattini_pd}.
If it is intuitively recognized that the presence of incomplete equilibration and 
collective flows may be treated in a statistical framework 
introducing extra constraints \cite{Rafelski}, 
the procedure is not necessarily unique.

The inclusion of collective motion in the form of a radial or elliptic 
flow in equilibrium models has been treated by different authors \cite{Dasgupta,Chikazumi,Richert,annals,shm,Pal,Samaddar}.
The most spread approach is to suppose a full decoupling between intrinsic and collective motion
and assume for the expanding system a standard Gibbs equilibrium in the local rest frame \cite{Bondorf,shm}. 
The quality of this assumption obviously depends on the degrees of freedom and energy regime under study. 
Concerning heavy ion collisions, this assumption may be justified 
in the Fermi energy regime because of the limited energy percentage associated to directed motion \cite{Wci}, 
and in the ultrarelativistic regime by the empirical success of hydrodynamical models \cite{hydro}. 
Some attempts have however been done to explicitly include flow in the statistical treatment.
Limiting ourselves to classical systems of interacting constituents treated as elementary degrees of freedom, 
the empirical treatment of flow in ref.\cite{Dasgupta} has been shown not 
to modify the correlation properties of the system.
However other empirical approaches \cite{Chikazumi,Richert} predict 
that the presence of flow should lead to a violation of 
statistical equilibrium weights, with a trend towards more unbound configurations. 
Experimental data in the nucleonic regime suggest that different mechanisms may act in distinct energy 
regimes \cite{Kunde,Reisdorf}.

%%%%%%%%%%%%%%%%%%%%%%%%%%%%%%%%%%%%%%%%%%%%%%%%%%%%%%%%%%%%%%%%%%%
%       INTRODUCTION: OUR INTENT
%%%%%%%%%%%%%%%%%%%%%%%%%%%%%%%%%%%%%%%%%%%%%%%%%%%%%%%%%%%%%%%%%%%
%
In this paper we address the generic statistical mechanics problem 
of the definition of a statistical ensemble in the presence of a collective flow.
We will use the example of a classical Lennard-Jones system \cite{Dorso}
to evaluate some chosen observables for a statistical 
isolated system subject to a radial flow. Molecular dynamics simulations on 
the same system have already shown 
that flow enhances partial energy fluctuations \cite{Ariel} and at the 
same time can act as a heat sink \cite{Chernomoretz,Matias}, 
cooling the system and thus preventing it to reach high temperatures. 
We will show that in the statistical limit it can also act as a heat bath, 
since the relaxation of the microcanonical constraint allows 
the isolated system to explore a larger configuration space.  
%
%%%%%%%%%%%%%%%%%%%%%%%%%%%%%%%%%%%%%%%%%%%%%%%%%%%%%%%%%%%%%%%%%%%
%       MODEL: FORMALISM
%%%%%%%%%%%%%%%%%%%%%%%%%%%%%%%%%%%%%%%%%%%%%%%%%%%%%%%%%%%%%%%%%%%
%
%
\section{ TIME DEPENDENT GIBBS ENSEMBLES}
In a recent paper \cite{annals} we have shown that flow naturally appears in the statistical
picture \cite{Jaynes,Balian} as soon as we introduce constraints
which are not constants of motion. 
Consider an isolated physical system characterized by a 
finite spatial extension $\langle R^2\rangle$
at a given time $t_0$. Introducing the density matrix 
$\hat{D}=\sum_{\left( n\right) }\left| \Psi ^{\left( n\right) }\right\rangle
\;p^{\left( n\right) }\;\left\langle \Psi ^{\left( n\right) }\right| $, the 
minimum biased microstate probability distribution $p^{(n)}$
is defined by  
\begin{equation}
\hat{D}_{\lambda_0}\left ( t_0 \right ) 
= \frac{1}{W_{\lambda_0}\left ( E \right )}\exp \left (-\lambda_0 \hat{R}^{2}\right ) 
\delta \left ( E - \hat{H}\right ),  \label{t0}
\end{equation}
where $\hat{H}$ is the Hamiltonian, $\lambda_0$ is a Lagrange multiplier constraining 
the finite size, and
\begin{equation}
W_{\lambda_0}\left ( E \right )=\sum_{(n)} \exp \left (-\lambda_0 {R}^{2}_{n}\right ) 
\delta \left ( E - {H}_{n}\right ) \label{entropy}
\end{equation}
is the associated density of states or partition sum.
The dynamical evolution of eq.(\ref{t0}) at times $t>t_0$ is obtained from the 
Liouville equation $\partial_{t}\hat{D} = -i/\hbar [ \hat{H},\hat{D} ]$ \cite{annals}, or 
equivalently from the time evolution of the constraint. In the Heisenberg representation
\begin{eqnarray}
\hat{R}^2(t)&=&e^{-i\Delta t \hat{H}} \hat{R}^2(t_0) e^{i\Delta t \hat{H}}\nonumber \\
&=& \hat{R}^2(t_0) + \sum_{p=1}^\infty \frac{(\Delta t)^p}{p!} \hat{B}^{(p)}, \label{evolution}
\end{eqnarray}
where $\Delta t=(t-t_0)$ and the $\hat{B}^{(p)}$ operators are defined by the recursive relation
\begin{equation}
\hat{B}^{(p)}=-\frac{i}{\hbar} \left [ \hat{H},\hat{B}^{(p-1)} \right ] \;\;\;;\;\;\; \hat{B}^{(0)}=\hat{R}^2. 
\end{equation}
The time dependence of the process can therefore be recasted in terms of an (a priori infinite)
number of extra constraints $\hat{B^{(p)}}$. In the simplified case of a system of non-interacting 
identical particles
\begin{equation}
\hat{H}=\sum_{i=1}^N \frac{\hat{p_i^2}}{2m}
\end{equation}
the series reduces to the two operators
\begin{eqnarray}
\hat{B}^{(1)} &=&-\frac{i}{\hbar} \left [\hat{H},\hat{R}^{2}\right ] =-\sum_{i=1}^N\frac{1}{m}%
\left( \hat{\vec{p}}_{i}\cdot \hat{\vec{r}}_{i}+\hat{\vec{r}}%
_{i}\cdot \hat{\vec{p}}_{i}\right) \label{B1}\\
\hat{B}^{(2)} &=&-\frac{i}{\hbar} \left [\hat{H},\hat{B}^{(1)}\right ]=\sum_{i=1}^N\frac{2\hat{p}_{i}^{2}}{m^{2}}. \label{B2}
\end{eqnarray}
Then the exact density matrix is given at any time $t>t_0$ by
\begin{eqnarray}
\hat{D}_{\lambda_0}(t)&=&\frac{\delta(E-\hat{H})}{W_{\lambda_0}(E,t)}\exp
\sum_{i=1}^N \Bigl[ -\beta\left( t\right) \frac{\hat{p}_{i}^{2}}{2m} 
\nonumber \\
&& 
-\lambda_0\hat{r}_{i}^{2}+\frac{\nu\left( t\right) }{2}
\left( \hat{\vec{p}}_{i}\cdot \hat{\vec{r}}_{i}+\hat{\vec{r}}%
_{i}\cdot \hat{\vec{p}}_{i}\right) \Bigr]  ,  \label{GP-expan}
\end{eqnarray}
with 
\begin{equation}
\beta \left( t\right) = \frac{2\lambda_0}{m}(\Delta t)^{2} \;\;\; , \;\;\; 
\nu   \left( t\right) = \frac{2\lambda_0}{m} \Delta t .
\label{beta-nu}
\end{equation}
The diabatic evolution of an isolated initially constrained freely expanding system
can then be described as a generalized Gibbs equilibrium in the local 
rest frame
\begin{equation}
\hat{D}_{\lambda_0}(t)=\frac{\delta(E-\hat{H})}{W_{\lambda_0}(E,t)}\exp
\sum_{i=1}^N-\beta\left( t\right) \frac{\left( \hat{\vec{p}}
_{i}-mh\left( t\right) \hat{\vec{r}}_{i}\right) ^{2}}{2m} \label{D-expan}
\end{equation}
with a Hubblian factor linearly decreasing in time, $h=\Delta t^{-1}$.

These equations show that radial flow is a necessary
ingredient of any statistical description of unconfined finite systems
in the presence of a continuum; on the other
hand, if a radial flow is observed in the experimental data, this formalism
allows to associate the flow observation to a
distribution at a former time when flow was absent. This initial
distribution corresponds to a static Gibbs equilibrium in a
confining harmonic potential.
In this case the infinite information which is a priori needed to follow the time evolution of the
density matrix according to eq.(\ref{evolution}), reduces to the three
observables $\hat{r}^{2}$, $\hat{p}^{2}$, $\hat{\vec{r}}
\cdot \hat{\vec{p}}+\hat{\vec{p}}\cdot \hat{\vec{r}}$. Indeed these
operators form a closed Lie algebra, and the exact 
evolution of $\hat{D}_{\lambda_0}$ preserves it algebraic structure.
This treatment can be easily extended to non-isotropic flows \cite{annals}
introducing an initially deformed spatial distribution.

It is easy to see that eq.(\ref{GP-expan}) is still exact for an interacting
system in the Boltzmann limit of purely local interactions.
If the interactions are non-local at the initial time $t_0$, this 
simple solution is not exact any more and higher order operators play a role. 
Considering a finite range two body interaction 
$\hat{V}=\sum_{ii^{\prime }}V(\left| \hat{\vec{r}}_{i}-\hat{\vec{r}}_{i^{\prime
}}\right|)$, we can see that the first order correction in time to the static
problem $\hat{B}^{(1)}$ is identical to the free problem eq.(\ref{B1}), 
while already at the second order $\hat{B}^{(2)}$ contains an additional term 
\[
\hat{B}^{(2)}=\sum_{i=1}^N2\frac{\hat{p}_{i}^{2}}{m^{2}}-\sum_{ii^{\prime }}%
\frac{1}{m}\hat{r}_{ii^{\prime }}\vec{\nabla}V(\hat{r}_{ii^{\prime}}) 
\]
where $\hat{r}_{ii^{\prime }}=\left| \hat{\vec{r}}_{i}-\hat{\vec{r}}_{i^{\prime}}\right| .$
In the case of a harmonic interaction the $\hat{B}^{(p)}$ operators only
contain quadratic terms $\sum_{i}\hat{p}_{i}^{2}$, $
\sum_{nn^{\prime }}\hat{r}_{ii^{\prime }}^{2}$ and $\sum_{ii^{\prime }}
\hat{r}_{ii^{\prime }}\cdot \hat{p}_{ii^{\prime }}$, with $
\hat{p}_{ii^{\prime }}=\hat{\vec{p}}_{i}-\hat{\vec{p}}
_{i^{\prime }}$. In this case the time evolution can be taken into account
by a suitable time dependent temperature and the
introduction of a radial flow. 

For any other interaction $\hat{B}^{(2)}$ 
modifies not only the temperature but
also the two-body interaction. As a first order approximation we can 
however still consider the statistical ansatz at the freeze-out time:
\begin{eqnarray}
W_{\tilde{\beta}\tilde{\lambda} \tilde{h}}&(E)& = \sum_{(n)} \exp \Bigl[ -\tilde{\beta} \sum_{i=1}^N \frac{1}{2m} 
\left( \vec{p}_{in} - \tilde{h} m \vec{r}_{in} \right)^2 \nonumber \\
&-& \beta V_{n} - \tilde{\lambda} R^{2}_{n}
\Bigr ] \delta\left ( H_{n}-E \right ), \label{basic}
\end{eqnarray}
where $\tilde{\beta},\tilde{\lambda},\tilde{h}$ are Lagrange parameters imposing 
a given value for the average thermal energy, mean square radius and local 
collective radial momentum at freeze out through the associated equations of state
\begin{eqnarray}
\langle E_{th} \rangle &=& -\frac{\partial W_{\tilde{\beta}\tilde{\lambda} \tilde{h}}}{\partial \tilde{\beta}} \\
\langle R^{2} \rangle &=& -\frac{\partial W_{\tilde{\beta}\tilde{\lambda} \tilde{h}}}{\partial \tilde{\lambda}} \\
\langle P_r\left (r\right ) \rangle &=& \frac{1}{\beta r}\frac{\partial W_{\tilde{\beta}\tilde{\lambda} \tilde{h}}}{\partial \tilde{h}} 
\end{eqnarray}
In heavy ion collisions, the values taken by these state variables are consequences of
the dynamics. They cannot be accessed by a statistical treatment but have to
be extracted from simulations and/or directly inferred from the data itself%\cite{cneg_exp}
.
In the following we take eq.(\ref{basic}) as an ansatz for the statistical description
of an expanding system and explore its properties within a classical system
of $N=147$ Lennard Jones particles of mass $m$ \cite{Dorso}.
We expect this ansatz to be reasonable in the case of loose interaction or 
moderate flows appearing at times close to the freeze-out time, and in the case 
of a fast re-organization of the potential energy surface, leading to a decoupling of 
the relaxation time of the interaction and kinetic energy.
The adequacy of eq.(\ref{basic}) to describe the time dependent expansion of the 
system will be explored in a forthcoming paper \cite{iguazu}.

The statistical ensemble described by eq.(\ref{basic}) is similar to a   
Gibbs equilibrium in the local expanding frame, with two important differences
with respect to the standard scenario \cite{Bondorf,shm} of a complete decoupling
between collective and thermal motion. 
First, the energy conservation constraint acts on the total energy, including flow.
This allows energy exchanges between the thermal and the collective motion, and 
therefore can modify considerably the partitions weight, as we show below.  
Second, eq.(\ref{basic}) contains a term $\propto r^{2}$ which plays the role 
of an external pressure \cite{Samaddar}.  
This term is the combination of a positive (out-going) pressure 
due to the expansion, and a negative pressure term imposing a finite system size 
at the freeze-out time. In turn this
implies that the correct ensemble for treating an open flowing system is 
not the usual $(N,V,T)$ or $(N,V,E)$ ensemble \cite{Gross,shm,Bondorf} but 
rather an ``isobar'' ensemble, where the system square radius is constrained 
only in average through a Lagrange parameter. 
This is an important point, since it is well known that different
statistical ensembles are not equivalent in finite systems \cite{noi,Barre}.
In particular only in such isobar ensemble the heat capacity is expected 
to be negative \cite{Gross} at the liquid-gas phase transition \cite{noi},
which is at the origin of an intense research in the nuclear multifragmentation
field \cite{cneg}. 
It is generally assumed by statistical models that fragment or hadron 
partitions are set within a characteristic volume (freeze-out volume) which
may depend on the thermal energy, but does not depend
on flow \cite{Bondorf,Papp,Sator,shm}. 
In this case the presence of flow does not affect the canonical 
configuration space of the isobar ensemble. Then flow can modify the partitions 
only because of the modified particle correlations 
in phase space \cite{Dasgupta, Mishustin, Richert,npa}, 
and because the microcanonical constraint 
acting on the total energy leads to a non trivial coupling between 
thermal and collective energy \cite{Dasgupta}.  
%
%
%
%
%%%%%%%%%%%%%%%%%%%%%%%%%%%%%%%%%%%%%%%%%%%%%%%%%%%%%%%%%%%%%%%%%%%
%       ANALYSIS
%%%%%%%%%%%%%%%%%%%%%%%%%%%%%%%%%%%%%%%%%%%%%%%%%%%%%%%%%%%%%%%%%%%
%
\section{SYSTEMS IN A HARMONIC TRAP} 
It is interesting to notice that eq.(\ref{basic}) is formally identical 
to a Gibbs equilibrium with an external harmonic potential 
$\hat{U}=\tilde{\lambda}/\tilde{\beta}\sum_i \hat{r}_i^2$.
The deep connection between an $\hat{R}^2$ constraint and radial collective 
motion is shown by the fact that it is extremely difficult from a technical 
point of view to equilibrate a Lennard Jones system in a harmonic trap; this
situation is referred to in the literature as ``the harmonic oscillator pathology''
 \cite{allen,harmonic}.
%
%   --- Figure 1
%
\begin{figure}[b!]
\setlength{\abovecaptionskip}{-10pt}
\begin{center}
\includegraphics[angle=-90, width=1\columnwidth, trim=1cm 0cm 0cm 0cm,clip=true]{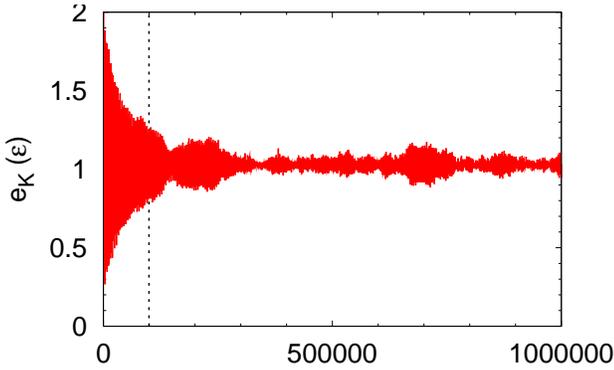}
\includegraphics[angle=-90, width=.95\columnwidth, trim=1cm 0cm 0cm 0cm,clip=true]{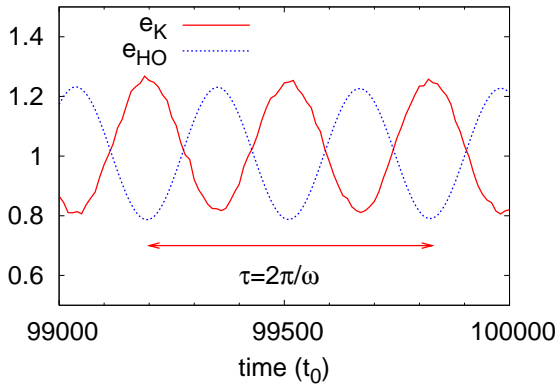}
\end{center}
\caption{\label{fig:fig1}
     (Color online) Upper part: total kinetic energy per particle as a function of time (in Lennard-Jones units) 
     for a Lennard-Jones system of 147 particles trapped in a harmonic oscillator 
     of spring constant $k$ such that $\omega=\sqrt{k/m}=0.01t_0^{-1}$ for a total energy per particle $e=2.0\epsilon$. 
     Lower part: zoom of the oscillation dynamics over a reduced time interval. 
     Dashed line: potential energy associated to the collective 
     oscillation $e_{HO}=\langle k\hat{r}^2/2\rangle$.
      }
\end{figure} 
\subsection{Dynamics of Lennard-Jones systems}

Figure 1 shows a single very long molecular dynamics run for the Lennard-Jones particles
trapped in a harmonic oscillator. Even if the amplitude of the initial oscillations is damped
by the inter-particle interaction, it is apparent from Figure 1 that collective oscillations
persist over extremely long times and the ergodic limit does not seem to be attained.
This situation is virtually independent of the chosen collective frequency
and total energy. As we now show, this behavior can be understood from the closed algebraic
structure of the $\hat{p}^2$, $\hat{r}^2$ and $\hat{\vec{r}}\cdot\hat{\vec{p}}$ operators, which is preserved 
in classical mechanics if commutators are replaced by Poisson brackets. 

Let us consider as above an initial condition given by eq.(\ref{t0})
within the ideal gas  $\hat{H}=\hat{E}_{K}+k\hat{R}^2/2$ or diluted Boltzmann limit.
If the only constraint on the size is given by the harmonic potential, the density matrix 
eq.(\ref{t0}) is a stationary solution of the Liouville equation.
If conversely the system is initialized to a different average size through
an extra constraint $\lambda_0\neq 0$, the system will evolve with the appearance of a 
collective flow $\hat{B}^{(1)}=-\sum_{n}\frac{1}{m}\left( \hat{\vec{p}}
_{n}\cdot \hat{\vec{r}}_{n}+\hat{\vec{r}}_{n}\cdot \hat{\vec{p}}
_{n}\right) $ as in eq.(\ref{B1}). Contrary to the free case, the successive constraining 
operators $\hat{B^{(p)}}$ do not vanish for any $p\geq 1$ and can be written as 
\begin{eqnarray}
\hat{B}^{(2p)}&=& \sum_{i=1}^N (-1)^p \left ( 2 \omega \right)^{2p} \left ( \frac{%
\hat{r}_{i}^{2}}{2}- \frac{\hat{p}_{i}^{2}}{2mk}\right ) \\
\hat{B}^{(2p+1)}&=& -\sum_{i=1}^N (-1)^p \left ( 2 \omega \right)^{2p} \frac{%
\hat{\vec{p}}_{i}\cdot \hat{\vec{r}}_{i} + \hat{\vec{r}}_{i} \cdot 
\hat{\vec{p}}_{i} }{m}
\end{eqnarray}
with $\omega=\sqrt{k/m}$. This gives at any time a density matrix with the same functional form 
as eq.(\ref{GP-expan}), with an effective temperature $\tilde{\beta}$, 
constraining field $\tilde{\lambda}$, and collective radial velocity $\tilde{\nu}$ oscillating in time.
For the purpose of getting analytical results it is easier to consider an initial condition in the canonical ensemble
\begin{equation}
\hat{D}_{\beta_0,\lambda_0}\left ( t_0 \right ) 
= \frac{1}{Z_{\beta_0,\lambda_0}}\exp \left (-\beta_0 \hat{H} - \lambda_0 \hat{R}^{2}\right ) .  \label{t0_can}
\end{equation}
The series eq.(\ref{evolution}) can be analytically summed up, and
the time dependent partition sum results $Z_{\tilde{\beta},\tilde{\lambda},\tilde{\nu} }= z_{\tilde{\beta},\tilde{\lambda},\tilde{\nu} }^N$ with
\begin{eqnarray}
z_{\tilde{\beta},\tilde{\lambda},\tilde{\nu} }(t)&=& 
Tr \Bigl[ \exp \Bigl( -\tilde{\beta} \left( t\right)  
\frac{\hat{p}^{2}}{2m} -\tilde{\lambda}\left( t\right) 
\hat{r}^{2} \nonumber \\
&+&\frac{\tilde{\nu}\left( t\right) }{2} \left ( \hat{\vec{p}}\cdot 
\hat{\vec{r}} + \hat{\vec{r}} \cdot \hat{\vec{p}}\right) \Bigr) \Bigr] 
, \label{breath}
\end{eqnarray}
The time dependent Lagrange parameters are given by
\begin{eqnarray}
\tilde{\beta}\left( t\right) &=& \beta_0 - \frac{\lambda_0}{k} \left ( \cos
2\omega \Delta t -1 \right )  \\
\tilde{\lambda}\left( t\right) &=& \frac{1}{2} \left ( \beta_0k + \lambda_0 \left  ( \cos 2 \omega \Delta t + 1\right ) \right ) \\
\tilde{\nu} \left( t\right) &=& \frac{\lambda_0}{m\omega} \sin 2\omega \Delta t. 
\end{eqnarray}
Eq.(\ref{breath}) can be interpreted as a Gibbs equilibrium in the rest frame of a
breathing system. For classical particles the trace over single-particle microstates is a phase-space integral $Tr[]= h^{-3} \int d^3r \int d^3p$ 
and the canonical partition sum is readily evaluated
\begin{equation}
z_{\tilde{\beta},\tilde{\lambda},\tilde{\nu} }(t)
=\frac{2\sqrt{\pi}m}{h^3} \left ( 2 \tilde{\beta} \tilde{\lambda} - 
\tilde{\nu}^2 m \right) ^{-3/2}
\end{equation}
This leads to the prediction for the time dependent behavior of the different
observables
\begin{eqnarray}
\frac{\langle p^2 \rangle}{2m} &=& -\frac{\partial \log z_{\tilde{\beta}\tilde{\lambda} \tilde{\nu}}}{\partial \tilde{\beta}}
= \frac{6 m \tilde{\lambda}}{2\tilde{\beta}\tilde{\lambda}-\tilde{\nu}^2m} \\
\langle r^{2} \rangle &=& -\frac{\partial \log z_{\tilde{\beta}\tilde{\lambda} \tilde{\nu}}}{\partial \tilde{\lambda}}
= \frac{3  \tilde{\beta}}{2\tilde{\beta}\tilde{\lambda}-\tilde{\nu}^2m} \\
\langle \vec{p}\cdot \vec{r} \rangle &=& 
\frac{\partial \log z_{\tilde{\beta}\tilde{\lambda} \tilde{\nu}}}{\partial \tilde{\nu}}
= \frac{3 m \tilde{\nu}}{2\tilde{\beta}\tilde{\lambda}-\tilde{\nu}^2m}
\end{eqnarray}
Introducing the expressions of $\tilde{\beta},\tilde{\lambda},\tilde{\nu}$
we get
\begin{eqnarray}
e_{K} \equiv \frac{\langle p^2 \rangle}{2m} &=&  
 \frac{3}{2\beta_0} 
\frac{1+x\left ( 1 + \cos 2 \omega \Delta t \right ) /2}{1+x} \\
e_{HO} \equiv \frac{k}{2}\langle r^{2} \rangle &=&  
 \frac{3}{2\beta_0} 
 \frac{1+x\left ( 1 - \cos 2 \omega \Delta t \right ) /2}{1+x} \\
e_{flow} \equiv \omega \langle \vec{p}\cdot \vec{r} \rangle &=& 
 \frac{3}{2\beta_0} \frac{x}{1+x} \sin 2 \omega \Delta t 
\end{eqnarray}
where $x=k_0/k$ measures the strength of the initial constraint.
It is clear from the inspection of Figure 1 that over
the time scale of a collective oscillation the interparticle interaction can be neglected, 
the total energy conservation constraint does not seem to play an important role, and the 
canonical free particles result eq.(\ref{breath}) appears fairly accurate.
The kinetic energy do oscillate with the double of the 
oscillator frequency in phase opposition, this collective motion breaking the ergodicity of the dynamics. 
%
%   --- Figure 2
%
\begin{figure}[hb]
\setlength{\abovecaptionskip}{0pt}
\begin{center}
\includegraphics[width=.9\columnwidth]{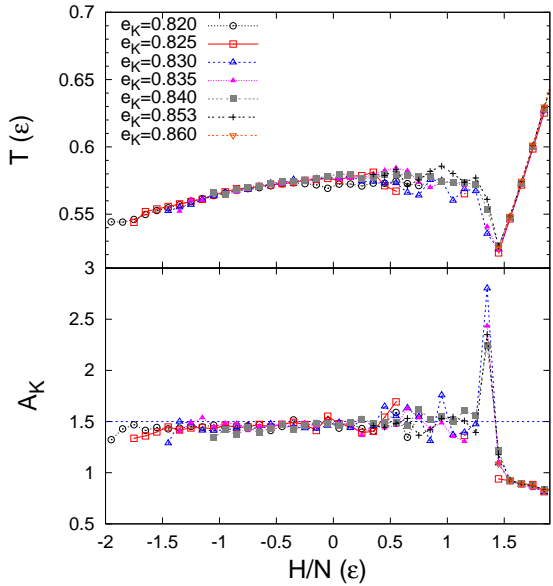}
\end{center}
\caption{\label{fig:fig2}
     (Color online) Microcanonical temperature (upper part) and normalized kinetic energy 
     fluctuation (lower part) as a function of the total energy inside 
     the harmonic oscillator obtained from an energy sorting of canonical 
     distributions corresponding to thermostat temperatures $\beta=3/(2 e_{K})$
     as indicated.
     All quantities are expressed in Lennard-Jones units.
     }
\end{figure}   
\subsection{Microcanonical Thermodynamics}
In order to study the effect of flow for the freely expanding system,
we have performed numerical molecular dynamics calculations within the 
statistical ensemble eq.(\ref{basic}) without ($h=0$) and with ($h>0$) 
the contribution of a radial collective flow.
To study the thermodynamical properties of the isobar ensemble characterized
by a size constraint $\lambda_0$ , we have 
constructed the microcanonical distribution by sorting a canonical ensemble \cite{Duflot}
of the equivalent system trapped in a harmonic oscillator of spring constant 
$k=2\lambda_0/\beta$.
The canonical distributions are obtained by coupling the system to a thermostat 
with the Andersen technique \cite{Andersen}.
In brief, the coupling is made by stochastic impulsive forces that act occasionally 
on randomly selected particles. After each collision, the selected particle is endowed 
with a new velocity drawn from a Maxwell-Boltzmann distribution at the desired 
temperature $T$. The combination of Newtonian dynamics with the stochastic collisions
 generate a Markov chain in phase space, which under some general conditions 
generates the canonical distribution \cite{Andersen}.

The resulting microcanonical thermodynamics is shown in Figure 2 
for an oscillator constant $\omega=0.01 t_0^{-1}$. Close to the liquid-gas 
transition temperature, the canonical calculations give rise to very wide 
energy distributions and the different events $(n)$ can be sorted in total energy bins
\begin{equation}
H_n= E_{Kn} + E_{LJn} +\frac{1}{2} k R^2_n \label{E_HO}
\end{equation}
This energy can be physically interpreted as a free enthalpy for the isolated unbound system
characterized by a finite size at the freeze-out time \cite{Duflot}. 
Each single canonical sampling can therefore be used to access the microcanonical thermodynamics
over a wide enthalpy region.
The microcanonical temperature is evaluated in each enthalpy bin as $T(H)=2<e_{K}>/3$ \cite{noi}
where the average is taken over events belonging to the same bin. The normalized kinetic energy 
fluctuation $A_K=N \left ( \langle e_{K}^2 \rangle - \langle e_{K} \rangle ^2\right ) /T^2$ is also 
represented. The nice agreement between estimations obtained with different canonical 
temperatures shows the quality of the numerical sampling. 
%
%   --- Figure 3
%
%\begin{figure}[t]
\begin{figure}[htb]
\setlength{\abovecaptionskip}{-10pt}  \centering
\unitlength1mm
\begin{picture}(80,80)
\put(-10.0,40.0)  {\parbox{3mm}{\psfig{figure=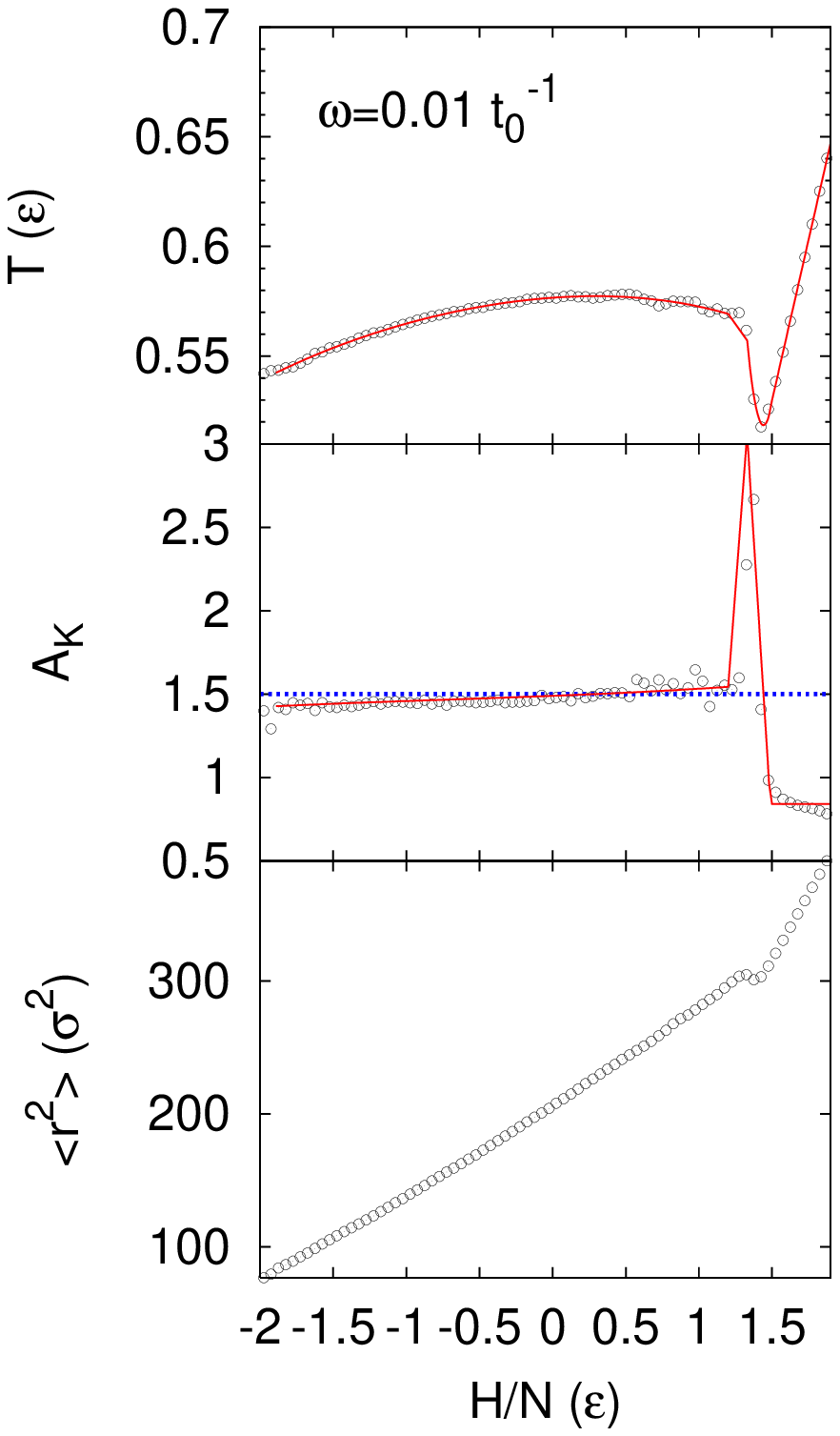,width=50mm}}}
\put(27.0,40.0) {\parbox{3mm}{\psfig{figure=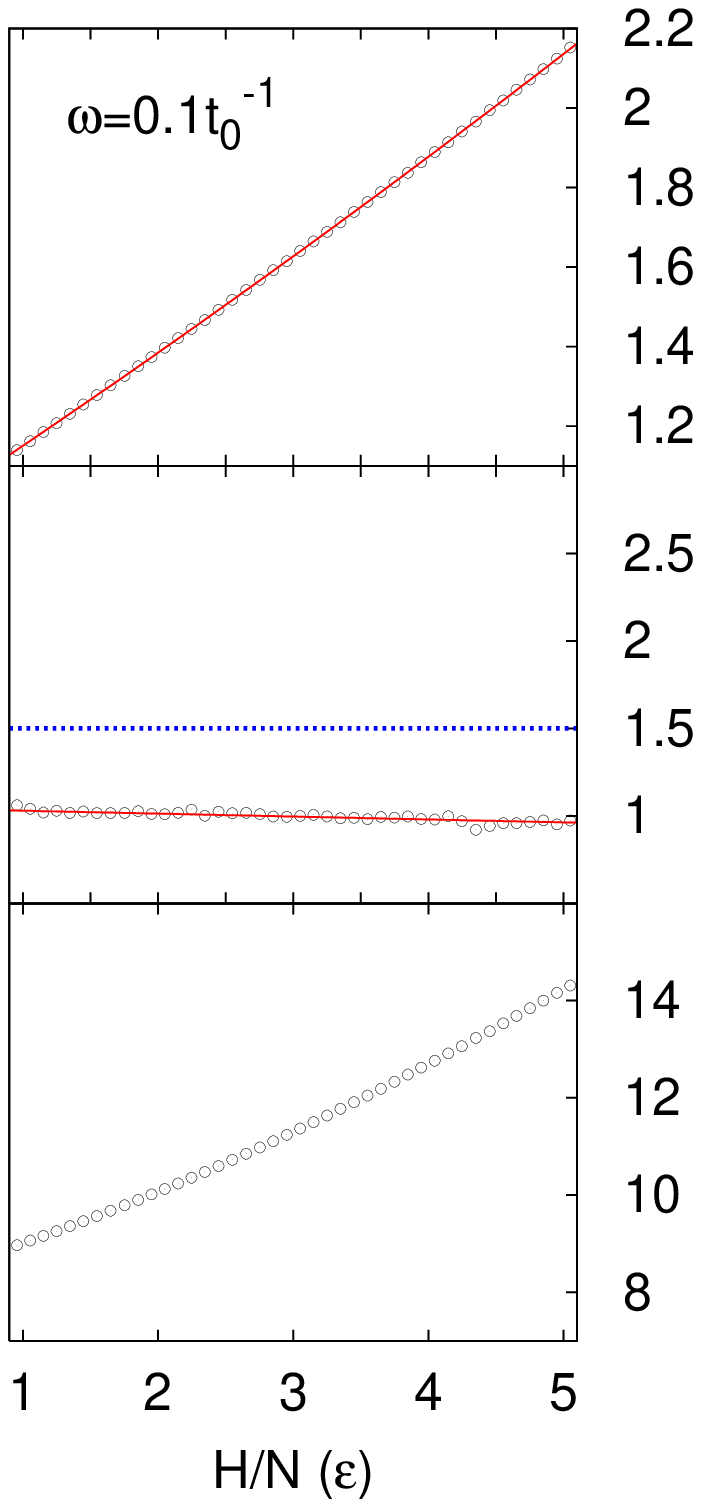,width=50mm}}}
\end{picture}
\caption{\label{fig:fig3}
     (Color online) Microcanonical temperature (upper part), normalized kinetic energy 
     fluctuation (medium part) and mean square radius (lower part) 
     as a function of the total energy inside 
     the harmonic oscillator for two different oscillator strengths.
     The horizontal lines in the medium panels give the fluctuation 
     expected in the canonical ensemble.
      }
\end{figure}
Figure 3 shows the dependence of the results on the oscillator strength.
We can recognize for low $\omega$ values, $i.e.$ loose constraints on the
system size, the first order liquid-gas phase transition. The transition 
is signaled by the backbending of the microcanonical caloric curve \cite{Gross}
corresponding to a negative heat capacity, and the associated abnormal 
kinetic energy fluctuation overcoming the canonical limit $A_K=3/2$ \cite{noi}.
The consistency between the two independent signals is again a proof of the 
numerical quality of the microcanonical sampling. 
These results are in qualitative agreement with the ones obtained for 
the Lattice Gas model in the same ensemble \cite{Duflot}.  
We can also notice that in the energy interval 
corresponding to the transition, the mean square radius shows a kink and a slope 
change at higher energies. The spatial extension of the unbound phase grows more rapidly
with the energy, and at the coexistence point the two phases have similar spatial
extensions. This means that in this model, contrary to the Lattice case \cite{noi},
the two coexisting phases at the transition temperature can be populated even in 
an ensemble which strongly constrains the volume of the system. In particular 
the two characteristic signals of a first order phase transition in a finite system,
namely bimodality in the canonical ensemble and negative heat capacity in the 
microcanonical one, can be observed even in the isochore ensemble \cite{Chernomoretz,Ariel}.   

For stronger size constraints (smaller average volumes) the caloric curve is monotonic, the 
microcanonical constraint reduces fluctuations well below the canonical limit, and 
the mean square radius increases linearly with the energy. This signals a supercritical system.
From these calculations the critical pressure can be roughly estimated as $\omega_c\approx 0.015 t_0^{-1}$. 

%We now turn to analyze the effect of flow on the thermodynamics of the system.
%To simulate the expanding ensemble eq.(\ref{basic}), a radial momentum 
%$\vec{p}_r=m\tilde{h}r\vec{u}_r$ is added to each particle and a microcanonical 
%sorting is imposed on the total energy including flow 
%$E=\sum_i (\vec{p}_i+\vec{p}_{ri})^2/(2m)$. The Hubble factor $\tilde{h}$ 
%employed at the different energies corresponds to the collective velocity
%obtained from the free expansion of the same system starting from a dense 
%configuration \cite{Chernomoretz}.
%Then calculations without flow at a total energy $E$ are compared to the ensemble
%including flow $h>0$ at a higher energy $E'=\langle E_{th} \rangle +\langle E_{flow}\rangle >E$, 
%such that the average thermal energy are similar, $\langle E_{th} \rangle=E$.

\section{MICROSTATE DISTRIBUTIONS IN AN EXPANDING ENSEMBLE}

To simulate the expanding ensemble eq.(\ref{basic}), a radial momentum
$\vec{p}_r=m\tilde{h}r\vec{u}_r$ is added to each particle and a microcanonical
sorting is imposed on the total energy including flow
$E'=\sum_i (\vec{p}_i+\vec{p}_{ri})^2/(2m) + E_{LJ}$. The Hubble factor $\tilde{h}$
employed at different energies has been obtained from the measured collective velocity
 of the same system freely expanding in vacuum ($E_{flow}^{free}$) according to
$h^2/2m <R^2>=E_{flow}^{free}$ \cite{Matias}.
Since the addition of flow trivially increases the total energy $E$, such that
$E'=\langle E_{th} \rangle +\langle E_{flow}\rangle >E$, the comparison between the
calculations without flow at an energy $E$ and those of the ensemble including flow
at an energy $E'$ have to be made such that the average thermal energy of both systems
are similar $\langle E_{th} \rangle=E$.

\begin{figure}[ht!]
\setlength{\abovecaptionskip}{-10pt}
\begin{center}
\includegraphics[width=.9\columnwidth,clip=]{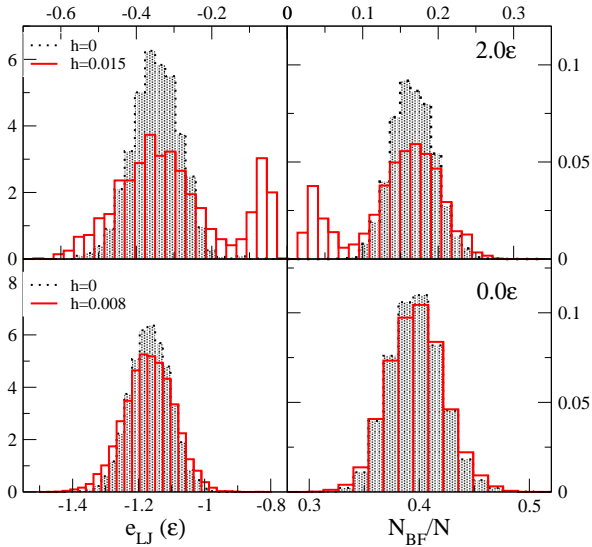}
\end{center}
%\begin{figure}[t]
%\unitlength1mm
%\begin{picture}(80,50)
%\put(-2.0,40.0)  {\parbox{3mm}{\psfig{figure=fig4.eps,width=40mm,trim=2.cm 0cm 0cm 0cm,clip=true}}}
%\put(40.0,40.0) {\parbox{3mm}{\psfig{figure=fig5.eps,width=41.5mm,trim=2.cm 0cm 0cm 0cm,clip=true}}}
%\end{picture}
\caption{\label{fig:fig4}
    (Color online) Potential energy (left side) and size of the largest cluster (right side) at two different
    thermal energies corresponding to the bound phase (lower part) and close to the transition
    region (upper part). The filled histograms correspond to a static equilibrium $h=0$ while the 
    empty ones flow was included according to eq.(\protect\ref{basic}). All quantities are expressed 
    in Lennard-Jones units.}
\end{figure}

The results are shown in Figure 4 for the distribution of the potential energy 
and the size of the largest fragment recognized through the MST algorithm \cite{Dorso}.
We can see that for all energies the presence of flow modifies the distributions 
in a sizeable way, leading to higher fluctuations.
This is easy to understand from eq.(\ref{basic}) if we consider that in the expansion 
dynamics only the total energy is conserved, meaning that thermal energy fluctuations
can be compensated by collective energy fluctuations.
In this sense the collective motion acts as a heat bath, leading to distributions 
similar to the canonical ones. 
In particular if the system has a total energy inside the coexistence region of the 
first order phase transition (upper part of Figure 4) the exchange with the flow reservoir 
can allow the system to explore the two coexisting phases. These latter differ in potential
energy $\Delta e_{LJ} \approx 0.4\epsilon$ but not in average spatial extension (see Figure 3)
and can therefore be accessed in the same ensemble 
for a given value of the average freeze-out volume.

This result implies that signals of phase transitions typical of the canonical ensemble 
such as bimodalities can be pertinent also in the microcanonical framework, if flow 
is accounted for in a thermodynamical consistent way. Then such signals
may be accessed even in experimental situations where the deposited energy 
is strongly constrained. A possible experimental confirmation of this prediction 
in nuclear multifragmentation can be found in ref.\cite{Pichon}.
 
At this point a word of caution is in order. Our ansatz (\ref{basic}) is exact only for a system
of non-interacting particles (or in the limit of local interactions). In the presence
of strong correlations this ansatz supposes the system relaxation time be small
compared to the time-scale of the expansion. This should be fulfilled if the average
collective velocity $\langle v_F\rangle$ is much smaller than the velocity associated 
to the thermal motion $\langle v_{th}\rangle$.
Within the ansatz (\ref{basic}) the local equation of state for the radial momentum reads
\begin{equation}
\langle p_r(r) \rangle \approx \frac{1}{r} \frac{\partial \log z}{\partial (\tilde{\beta}\tilde{h})}
=\tilde{h}m r
\end{equation}
where we have neglected the effect of the energy-conserving $\delta$-function 
in eq.(\ref{basic}) in order to have an analytical order-of-magnitude estimate.
This leads to a collective velocity $v_F=\tilde{h}\langle R \rangle$ which should be 
compared to the canonical estimate $v_{th}=\sqrt{3/(\tilde{\beta}m)}$.
In the case of the upper part of Figure 4 we have $v_F/v_{th}\approx 0.79$ 
meaning that the quality of our approximation may be doubtful.
It is however interesting to note that the bimodal shape of the distribution in the presence of flow persists also for smaller collective motions, as long as the energy fluctuations are of the order of the energy distance $\Delta e_{LJ}$ 
between the two phases. 

%In the case of the Lennard-Jones systems we have $\Delta V\approx 0.4\epsilon$ which is comparable to the typical fluctuations for the kinetic terms. Indeed the thermal kinetic energy fluctuation in the canonical approximation is $\Delta K= \sqrt{\sigma^2_K}=\sqrt{2/3N}\langle K\rangle\approx 0.16\epsilon$ while the flow fluctuation can be evaluated through the relation
%
%\begin{equation}
%\sigma^2_{P_r}= \frac{\partial^2 \log Z}{\partial (\tilde{\beta}\tilde{h})^2}
%=\frac{m}{\tilde{\beta}}
%\end{equation}
%
%leading for the upper part of Figure 4 to 
%$\Delta E_F= (2\tilde{\beta})^{-1}=\langle K \rangle /3\approx 0.8 \epsilon$. 
%This order-of-magnitude estimate means that the scenario we have proposed may have some relevance in the phenomenology of the liquid-gas phase transition.

%
%%%%%%%%%%%%%%%%%%%%%%%%%%%%%%%%%%%%%%%%%%%%%%%%%%%%%%%%%%%%%%%%%%%%%%%%%%%%%%%%%%
% CONCLUSIONS
%%%%%%%%%%%%%%%%%%%%%%%%%%%%%%%%%%%%%%%%%%%%%%%%%%%%%%%%%%%%%%%%%%%%%%%%%%%%%%%%%%
%
\section{CONCLUSIONS}

To conclude, in this paper we have presented an information theory based formalism
allowing to include collective motions in the statistical description 
of finite bound or unbound systems. Molecular dynamics simulations 
performed on a Lennard-Jones system suggest that even in the simplified 
approximation of non interacting particles, the presence of flow can influence 
the microstate distribution in a sizeable way.
Indeed the presence of a (non-conserved in time) collective energy component
can play the role of a heat bath, allowing for extra configurational 
energy fluctuations in the total energy conserving dynamics. 
In particular, close to a first order phase transition, this mechanism is seen
to give rise to a characteristic bimodal behavior, similar to some recent 
experimental observations in nuclear multifragmentation.
%  
%
%%%%%%%%%%%%%%%%%%%%%%%%%%%%%%%%%%%%%%%%%%%%%%%%%%%%%%%%%%%%%%%%%%%%%%%%%%%%%%%%%%
% ACKNOWLEDGEMENTS
%%%%%%%%%%%%%%%%%%%%%%%%%%%%%%%%%%%%%%%%%%%%%%%%%%%%%%%%%%%%%%%%%%%%%%%%%%%%%%%%%%
%\bigskip
\section*{ACKNOWLEDGEMENTS}
This work was partially supported by the University of Buenos Aires via Grant X360. 
M.J.I. acknowledges the warm hospitality of the Laboratoire de Physique Corpusculaire 
(LPC) at Caen, and financial support from the LPC, the University of Buenos Aires, and 
Fundacion Antorchas.

%

%%%%%%%%%%%%%%%%%%%%%%%%%%%%%%%%%%%%%%%%%%%%%%%%%%%%%%%%%%%%%%%%%%%
%
%       T H E       E N D
%
%%%%%%%%%%%%%%%%%%%%%%%%%%%%%%%%%%%%%%%%%%%%%%%%%%%%%%%%%%%%%%%%%%%
%
\end{document}